\documentclass[twocolumn]{IEEEtran} 
\makeatletter
\let\NAT@parse\undefined
\makeatother

\usepackage{amsmath,amssymb,amsfonts,cuted}
\usepackage[final]{graphicx}
\usepackage{psfrag}
\usepackage{epsfig}
\usepackage[numbers,sort&compress]{natbib}

\usepackage{array,booktabs}

\usepackage{amsmath,amssymb,amsfonts,amsthm}
\usepackage[final]{graphicx}

\usepackage[nolist,nohyperlinks]{acronym}

\usepackage{ucs} 
\usepackage[utf8x]{inputenc}
\usepackage[usenames,dvipsnames]{xcolor}
\usepackage{tikz}
\usepackage{tkz-tab}
\usepackage{multirow}
\usepackage{latexsym}
\usepackage{amssymb}
\usepackage{amsmath}
\usepackage{mathrsfs}
\usepackage{amsthm}

\begin{acronym}
\acro{SNR}{signal-to-noise ratio}
\acrodefplural{SNR}[SNRs]{signal-to-noise ratios}
\acro{AWGN}{additive white Gaussian noise}
\acro{CSI}{channel state information}
\acro{PDF}{probability density function}
\acrodefplural{PDF}[PDFs]{probability density functions}
\acro{CDF}{cumulative distribution function}
\acrodefplural{CDF}[CDFs]{cumulative distribution functions}
\acro{RV}{random variable}
\acro{LoS}{line-of-sight}
\acro{MRT}{maximal ratio transmission}
\acro{MRC}{maximal ratio combining}
\acro{dR}{double-Rayleigh}
\acro{CLT}{central limit theorem}
\acro{i.i.d.}{independent and identically distributed}
\acro{fdRLoS}{fluctuating double-Rayleigh with line-of-sight}
\acro{dRLoS}{double-Rayleigh with line-of-sight}
\acro{SOSF}{second order scattering fading}
\acro{OP}{outage probability}
\acro{MC}{Monte Carlo}
\end{acronym}

\begin{document}

\title{\LARGE{A Fluctuating Line-of-Sight Fading Model \\with Double-Rayleigh Diffuse Scattering}}
\author{Jes\'us L\'opez-Fern\'andez, Pablo Ram\'irez Espinosa, Juan M. Romero-Jerez and F.~Javier L\'opez-Mart\'inez

\thanks{This work has been funded by the Spanish Government and the European Fund for Regional Development FEDER (project TEC2017-87913-R), and by Junta de Andaluc\'ia (project P18-RT-3175).}


\thanks{J. L\'opez-Fern\'andez, F. J. L\'opez-Mart\'inez and J. M. Romero-Jerez are with Communications and Signal Processing Lab, Instituto Universitario de Investigaci\'on en Telecomunicaci\'on (TELMA), Universidad de M\'alaga, CEI Andaluc\'ia TECH, ETSI Telecomunicaci\'on, Bulevar Louis Pasteur 35, 29010 M\'alaga (Spain). (contact e-mail: $\rm jlf@ic.uma.es$).}

\thanks{\noindent P. Ram\'irez-Espinosa is with the Connectivity Section,  Department of Electronic Systems, Aalborg University, Aalborg {\O}st 9220, Denmark.}

}
\maketitle

\begin{abstract}
We introduce the \ac*{fdRLoS} fading model as a natural generalization of the double-Rayleigh with line-of-sight fading model, on which the constant-amplitude line-of-sight component is now allowed to randomly fluctuate. We discuss the key benefits of the {fdRLoS} fading model here formulated over the state of the art, and provide an analytical characterization of its chief probability functions. We analyze the effect of the fading parameters that define the model, and discuss their impact on the performance of wireless communication systems.
\end{abstract}


\vspace{-3mm}
\section{Introduction}
The ubiquity of wireless systems has been enabled by the combination of multipath propagation, reflection and diffraction that makes communications possible without the existence of a \ac{LoS} between the transmitter and receiver ends. However, in many circumstances the existence of \ac{LoS} is an inherent characteristic of the propagation conditions in a number of environments, especially as wireless systems move toward higher frequencies \cite{Zochmann2019,Reig2019}.

The most popular fading model for \ac{LoS} propagation is the classical Rician model \cite{Rice1945}. It is based on the \ac{CLT}, by assuming that a sufficiently large number of multipath components arrive at the receiver end. The direct path is identified with the \ac{LoS} component and the aggregation of the remaining  waves is approximately Gaussian. There are different ways to generalize Rician fading model in order to allow it to model a wider range of propagation conditions. One of such generalizations is the so-called \ac{SOSF} model introduced in \cite{Andersen2002} and later formalized in \cite{Salo2006}, on which an additional term associated to double-scattering was introduced. A special case of this \ac{SOSF} model is the \ac{dRLoS} fading model, on which the Gaussian diffuse component that arises from the \ac{CLT} is not considered and the overall scattering corresponds to double-scattering, which in general implies a larger fading severity than its Rayleigh counterpart \cite{Lopez2018}. Such propagation conditions occur in the context of backscatter communications \cite{Devineni2019}, or in the context of optical wireless communications through the so-called I-K distribution \cite{Andrews1985}. Further generalizations to incorporate multiple scattering components of increasing order are available in the literature \cite{Salo2006,Nikolaidis2018}, at the expense of a much more sophisticated mathematical formulation.

Another popular generalization of the Rician fading model was introduced by Adbi et al. in \cite{Abdi2003}, by allowing the \ac{LoS} component to randomly fluctuate. This model, usually referred to as Rician shadowed fading model, has a number of appealing properties: for instance, it does not only generalize Rician fading model by adding an additional parameter that models a more general propagation condition, but in some cases, its mathematical representation is even simpler than the original Rician fading model \cite{Paris2010}.

The goal of this paper is to introduce a different generalization of Rician fading that includes both \ac{LoS} fluctuation and second-order scattering at the same time: we refer to this distribution as the \ac{fdRLoS} fading model. We will see how the \mbox{consideration} of the random fluctuations in the \ac{LoS} components allow to alleviate some of the limitations of the baseline \ac{dRLoS} fading model, such as its peaky behavior that is not representative to real data which causes an overestimation of the peak probability density \cite{Churnside1989}. We will also analyze how the performance of wireless communication systems operating over \ac{fdRLoS} fading channels is affected by the different propagation conditions captured by the model.

\textit{Notation}: $\mathbb{E}\{X\}$ and $|X|$ denote the statistical average and the modulus of the complex \ac{RV} $X$ respectively. The \ac{RV} $X$ conditioned to $Y$ will be denoted as $X|Y$. The symbol $\sim$ reads as \emph{statistically distributed as}. The symbol $\stackrel{d}{=}$ reads as \emph{equal in distribution}. A circularly symmetric normal \ac{RV} $X$ with mean $\mu$ and variance $\Omega$ is denoted as $X\sim \mathcal{N}_c(\mu,\Omega)$.
%
\vspace{-2mm}
\section{Physical models}
For the sake of comprehensiveness, and in order to allow for a better understanding of the relationship between the \ac{fdRLoS} fading model with other state-of-the-art alternatives, we briefly detail in the sequel the physical models for the most relevant amplitude-based \ac{LoS} fading models in the literature.
\label{Sec:The system model}
%
%
\vspace{-2mm}
\subsection{Physical model for Rician fading}
The physical model for the received signal $S$ under the Rician fading model \cite{Rice1945,Simon2005} is given by
\begin {equation}
S=\omega_0 e^{j\phi}+\omega_1 G_1,
\label{Eq:Modelo_Rician}
\end{equation}
where $\omega_0 e^{j\phi}$ is the \ac{LoS} component with average amplitude $\omega_0$, $\phi$ is a \ac{RV} uniformly distributed in $[0,2\pi)$, and $G_1$ is a zero-mean normal complex Gaussian \ac{RV}, i.e. $G_1\sim\mathcal{N}_c(0,1)$. The weighting factor $\omega_1$ is real and non-negative, and scales the variance of $G_1$. Besides its average power $\Omega=\mathbb{E}\{S^2\}=\omega_0^2+\omega_1^2$, the Rician model is completely defined by the so-called Rician $K$ parameter, defined as
\begin{equation}
\label{eq:K}
K=\tfrac{\omega_0^2}{\omega_1^2}\,\in\,[0,\infty),
\end{equation}
which accounts for the ratio between the powers associated to the \ac{LoS} and non-\ac{LoS} components.
\vspace{-2mm}
\subsection{Physical model for Rician shadowed fading}
The physical model for the received signal $S$ under the Rician shadowed fading model \cite{Abdi2003} is given by
\begin {equation}
S=\omega_0 \sqrt{\xi} e^{j\phi}+\omega_1 G_1,
\label{Eq:Modelo_RS}
\end{equation}
where the parameters in \eqref{Eq:Modelo_RS} are equivalent to those in \eqref{Eq:Modelo_Rician}, and $\xi$ is a Gamma distributed \ac{RV} with unit power and real positive shape parameter $m$, with \ac{PDF}:
\begin{equation}
f_{\xi}(u)=\tfrac{m^mu^{m-1}}{\Gamma(m)}e^{-m u},
\end{equation}
where $\Gamma(\cdot)$ is the gamma function. This variable $\xi$ captures the \ac{LoS} fluctuations, ranging from very severe for low $m$ to milder ones as $m$ grows. In the limit case of $m\rightarrow\infty$, $\xi$ degenerates to a deterministic unitary value and the \ac{LoS} fluctuation vanishes, then specializing to the conventional Rician fading model.

The Rician shadowed model is characterized by two parameters: $K$, as in \eqref{eq:K}, and $m\in(0,\infty)$.
\vspace{-2mm}
\subsection{Physical model for \ac{dRLoS} fading}
The physical model for the received signal $S$ under \ac{dRLoS} fading is expressed as in \cite{Salo2006}:
\begin {equation}
S=\omega_0 e^{j\phi}+\omega_2 G_2 G_3,
\label{Eq:Modelo_dRLoS}
\end{equation}
where $\omega_0$ and $\phi$ inherit the same definitions as in the Rician model, $G_2$ and $G_3$ are \ac{i.i.d.} zero-mean normal complex Gaussian \ac{RV}s, i.e. $G_i\sim\mathcal{N}_c(0,1)$ for $i=2,3$. The component $\omega_2 G_2 G_3$ leads to a \ac{dR} fading component, and hence, it differs from classical Rician fading for which the diffuse component lead to a single-Rayleigh \ac{RV}.

The parameters $\omega_0$ and $\omega_2$ are restricted to being real and non-negative. As in the Rician fading model, the parameter \color{black} $K$ suffices to completely define the distribution, i.e.,
\begin{equation}
\label{eq:K_dRLos}
K=\tfrac{\omega_0^2}{\omega_2^2}\,\in\,[0,\infty),
\end{equation}
The model in (\ref{Eq:Modelo_dRLoS}) includes the \ac{dR} fading model when $K=0$.\color{black}
\vspace{-2mm}
\subsection{Physical model for fluctuating \ac{dRLoS} fading}
The physical model for the received signal $S$ under the \ac{dRLoS} fading model with \ac{LoS} fluctuations is a natural generalization of the \ac{dRLoS} model introduced by Salo et al. \cite{Salo2006}, as
\begin {equation}
S=\omega_0 \sqrt{\xi} e^{j\phi}+\omega_2 G_2 G_3,
\label{Eq:Modelo_fdRLoS}
\end{equation}
where $\omega_0 e^{j\phi}$ is the \ac{LoS} component with constant amplitude $\omega_0$, $\phi$ is a \ac{RV} uniformly distributed in $[0,2\pi)$, $G_2$ and $G_3$ are \ac{i.i.d.} zero-mean normal complex Gaussian \ac{RV}s, i.e. $G_i$ is distributed as $\mathcal{N}_c(0,1)$ for $i=2,3$, and $\xi$ is a Gamma distributed \ac{RV} with unit power and real positive shape parameter $m$. 
As in the Rician shadowed fading model, the \ac{fdRLoS} fading model is completely defined by two parameters {\color{black}$K$} as in \eqref{eq:K_dRLos} and $m$.%

\vspace{-2mm}
\section{Statistical Characterization}
\label{Sec:3}
In this section, we will derive analytical expressions for the \ac{PDF} and the \ac{CDF} of the instantaneous \ac{SNR} $\gamma$ under \ac{fdRLoS} fading. Assuming a normalized channel with $\mathbb{E}\{|S|^2\}=1$, we have that $\gamma=\overline\gamma|S|^2$, where $\overline\gamma$ is the average \ac{SNR}. We will base our derivations on the key findings detailed in \cite{Lopez2018}, that will allow us to connect the \ac{fdRLoS} fading distribution with an underlying Rician shadowed distribution.

Let us begin by considering \eqref{Eq:Modelo_fdRLoS}, so that we have
\begin {equation}
\gamma=\overline\gamma|\omega_0 {\color{black}\sqrt{\xi}}e^{j\phi}+\omega_2 G_2 G_3|^2.
\label{Eq:gamma}
\end{equation}
Now, we express the complex Gaussian RV $G_3$ as $G_3=|G_3|e^{j\Psi}$, where $\Psi$ is uniformly distributed in $[0,2\pi)$. Because $G_2$ is a circularly-symmetric \ac{RV}, the following equivalence in distribution holds for $\gamma$ 
\begin {equation}
\gamma\stackrel{d}{=}\overline\gamma|\omega_0 {\color{black}\sqrt{\xi}}e^{j\phi}+\omega_2 G_2| G_3||^2.
\label{Eq:gamma2}
\end{equation}
Conditioning on $x=|G_3|^2$, define the conditioned \ac{RV} $\gamma_x$ as
\begin {equation}
\gamma_x\triangleq\overline\gamma|\underbrace{\omega_0 {\color{black}\sqrt{\xi}}e^{j\phi}+\omega_2 \sqrt{x} G_2}_{S_x}|^2.
\label{Eq:gamma3}
\end{equation}
We see that $S_x$ is a Rician shadowed \ac{RV} as in \eqref{Eq:Modelo_RS}, with $\omega_1=\omega_2 \sqrt{x}$. Hence, we have that $\gamma_x$ is distributed according to a squared Rician {\color{black} shadowed} \ac{RV} with parameters $m$,
\begin{align}
\overline\gamma_x&=\mathbb{E}\{\gamma_x\}=\overline\gamma(\omega_0^2+x\omega_2^2)=\overline\gamma\tfrac{K+x}{K+1},\\
K_x&=\tfrac{\omega_0^2}{x\omega_2^2}=\tfrac{K}{x},
\end{align}
where $K=\omega_0^2/\omega_2^2$ is the Rician factor in the absence of the \ac{RV} $x$, i.e., as in \eqref{eq:K_dRLos}. Hence, the \ac{PDF} of $\gamma_x$ is that of the \ac{SNR} of a Rician shadowed \ac{RV}, i.e. \cite{Abdi2003,Paris2014},
\begin{equation}
f_{\gamma_x}(\gamma;x)=\tfrac{m^m(1+K_x)}{(m+K_x)^m\overline\gamma_x}e^{-\tfrac{1+K_x}{\overline\gamma_x}\gamma}{}_1F_{1}\left(m;1;\tfrac{K_x(1+K_x)}{K_x+m}\tfrac{\gamma}{\overline\gamma_x}\right),
\label{eqaux}
\end{equation}
where $_1F_{1}(\cdot;\cdot;\cdot)$ denotes the Kummer confluent hypergeometric function \cite[eq. (16.2)]{NIST}. Noting that $|G_2|^2$ is exponentially distributed with unitary mean, we can compute the distribution of $\gamma$ by averaging over all possible values of $x$ as:
\begin{equation}
f_{\gamma}(\gamma)=\int_0^\infty f_{\gamma_x}(\gamma;x) e^{-x}dx.
\label{Eq_int_PDF}
\end{equation}
Plugging \eqref{eqaux} into \eqref{Eq_int_PDF}, an integral expression for the PDF of the \ac{fdRLoS} model is derived. In order to obtain a simpler expression for $f_{\gamma}(\gamma)$, we consider that the parameter $m\in\mathbb{Z}^+$. We note that such restriction does not cause a major impact unless the \ac{LoS} fluctuation is very severe; as we will later see, the practical benefits of the \ac{fdRLoS} fading model will become apparent precisely for mild and medium fluctuations of the \ac{LoS} component. Under this premise, the \ac{PDF} of the Rician shadowed fading model simplifies as \cite[eq. (5)]{Paris2014}
\begin{equation}
f_{\gamma_x}(\gamma;x)=\sum_{j=0}^{m-1}\tfrac{C_j(x)}{(m-j{\color{black}-1})!}\tfrac{\gamma^{m-j{\color{black}-1}}}{\Omega(x)^{m-j}}e^{-\tfrac{\gamma}{\Omega(x)}},
\label{Eq_PDF_RS}
\end{equation}
with $C_j(x)=\binom{m{\color{black}-1}}{j}\left(\tfrac{mx}{mx+K}\right)^j \left(\tfrac{K}{K+mx}\right)^{m-1-j}$ and $\Omega(x)=\overline\gamma\tfrac{K+mx}{m(K+1)}
$.
 {\color{black} Substituting (\ref{Eq_PDF_RS}) in (\ref{Eq_int_PDF}), using the change of variables ${t=\tfrac{1}{m}(K+mx)}$ and taking into account that ${(t-K/m)^j=\sum_{r=0}^{j}\binom{j}{r}t^j (-K/m)^{j-r}}$, the following expression for the \mbox{fdRLoS} PDF is derived:

 \begin{align}
 \label{eqPDFfdRLoS}
 &f_{\gamma}(\gamma)=\sum_{j=0}^{m-1} \binom{m-1}{j}\tfrac{(K/m)^{m-j-1}(K+1)^{m-j}e^{K/m}}{\overline{\gamma}^{m-j}(m-j-1)!} \gamma^{m-j-1} \times \nonumber \\
&\sum_{r=0}^{j} \binom{j}{r} \left(\tfrac{-K}{m} \right)^{j-r} \Gamma\left( \small{r+j-2m+2},\tfrac{K}{m},\tfrac{\gamma}{\overline \gamma}(K+1)\right),
 \end{align}

}
where $\Gamma(a,z,b)=\int_{z}^{\infty}t^{a-1} e^{-t}e^{\tfrac{-b}{t}}dt$ is a generalization of the incomplete gamma function defined in \cite{CHAUDHRY199499}. 
 
\begin{figure}[t]
\centering
\includegraphics[width=.78\columnwidth]{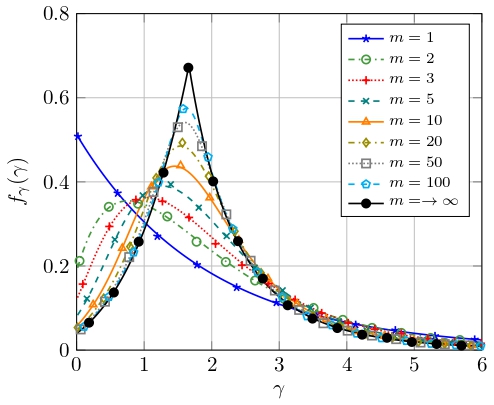}
\centering
\caption{PDF of the \ac{fdRLoS} fading model for different values of $m$ as a function of $\gamma$ (in linear form). Parameter values are $K=5$ and $\overline\gamma_{\rm dB}=3$dB. Theoretical values \eqref{eqPDFfdRLoS} are represented with lines. Markers correspond to \ac{MC} simulations. The case $m\rightarrow\infty$ is the \ac{dRLoS} \ac{PDF} in \cite[eq. (15)]{Lopez2018}.}
\label{Fig1}
\end{figure}

The \ac{CDF} of the \ac{fdRLoS} fading model can also be obtained by averaging the \ac{CDF} of $\gamma_{x}$, i.e., the Rician shadowed \ac{CDF} over the exponential distribution:
\begin{equation}
\label{CDF_promediado}
F_{\gamma}(\gamma)=\int_0^\infty F_{\gamma_x}(\gamma;x) e^{-x}dx.
\end{equation}
\color {black}
For the case of integer $m$, a closed-form expression for the Rician shadowed CDF is presented in \cite[eq. (10)]{Paris2014}, i.e. 
\begin{equation}
\label{CDF_Rice_Shadowed}
F_{\gamma_x}(\gamma;x)=1-\sum_{j=0}^{m-1}C_j(x)e^{-x/\Omega(x)} \sum_{r=0}^{m-j-1}\tfrac{1}{r!}\left( \tfrac{x}{\Omega(x)}\right)^r,
\end{equation}
Plugging (\ref{CDF_Rice_Shadowed}) into (\ref{CDF_promediado}), the following expression for the \ac{fdRLoS} CDF is obtained:
 \begin{align}
 \label{eqCDFfdRLoS}
 &F_{\gamma}(\gamma)=1-\sum_{j=0}^{m-1} \binom{m-1}{j} \left(\tfrac{K}{m} \right)^{m-j-1}e^{K/m} \times \nonumber \\
&\sum_{r=0}^{m-j-1}\tfrac{1}{r!}\left( \tfrac{\gamma}{\overline \gamma}\right)^r (K+1)^r  \times \\
&\sum_{s=0}^{m-j-1} \binom{j}{s}\left(\tfrac{-K}{m}\right)^{j-s} \Gamma\left(s-m-r+2,\tfrac{K}{m},\tfrac{\gamma}{\overline \gamma}(K+1)\right). \nonumber
 \end{align}
\color {black}
In Fig. \ref{Fig1}, we represent the evolution of the \ac{PDF} of the \ac{fdRLoS} fading model using \eqref{eqPDFfdRLoS} as the \ac{LoS} fluctuation severity $m$ changes, for $K=5$ and $\overline\gamma_{\rm dB}=3$dB (i.e., $\overline\gamma=2$). \ac{MC} simulations are included in all instances to double-check the corresponding theoretical expressions. We see that as $m$ is increased, i.e., the fading severity of the \ac{LoS} component is decreased, the \ac{SNR} values are less disperse and lower \ac{SNR} values are less likely. We see that for large $m$, the \ac{PDF} of the \ac{fdRLoS} fading model tends to behave as the \ac{PDF} of the \ac{dRLoS} fading model (in black solid line); however, note that the peaky behavior exhibited by the \ac{dRLoS} fading model does not appear in the \ac{fdRLoS} case, which has a smoother shape.
 
%

\vspace{-2mm}
\section{Application: Outage Probability}

\color{black} The \ac{OP} is defined as the probability that the instantaneous SNR takes a value below a given threshold $\gamma_{\rm th}$, which is directly computed from the CDF in (\ref{eqCDFfdRLoS}) as $\text{OP}=F_{\gamma}(\gamma_{\rm th})$. \color{black}
In the high-\ac{SNR} regime, an asymptotic expression for the \ac{OP} can be obtained as follows: since the asymptotic \ac{OP} under Rician shadowed fading is given by \cite{Paris2014}
\begin{equation}
{\rm OP_{RS}}\left(\gamma_{\rm th};\overline\gamma,K,m\right)=t\frac{\gamma_{\rm th}}{\overline\gamma}(1+K)\left(\tfrac{m}{K+m}\right)^m,
\end{equation}
then using the same rationale as in Section \ref{Sec:3} we have that
\begin{align}
{\rm OP}\left(\gamma_{\rm th};\overline\gamma,K,m\right)&=\int_0^{\infty}{\rm OP_{RS}}\left(\gamma_{\rm th};\overline\gamma(x),K(x),m\right)e^{-x}dx,\notag\\
&=\tfrac{\gamma_{\rm th}}{\overline\gamma}\underbrace{(1+K)\Gamma(m) {\mathrm U}\left(m,1,K/m\right)}_{a},\label{aOP}
\end{align}
where $\overline\gamma(x)=\overline\gamma\frac{K+k}{K+1}$ and $K(x)=K/x$, and ${\mathrm U}\left(\cdot,\cdot,\cdot \right)$ is Tricomi's confluent hypergeometric function \cite[(13.1)]{NIST}. 

\begin{figure}[t]
\centering
\includegraphics[width=.78\columnwidth]{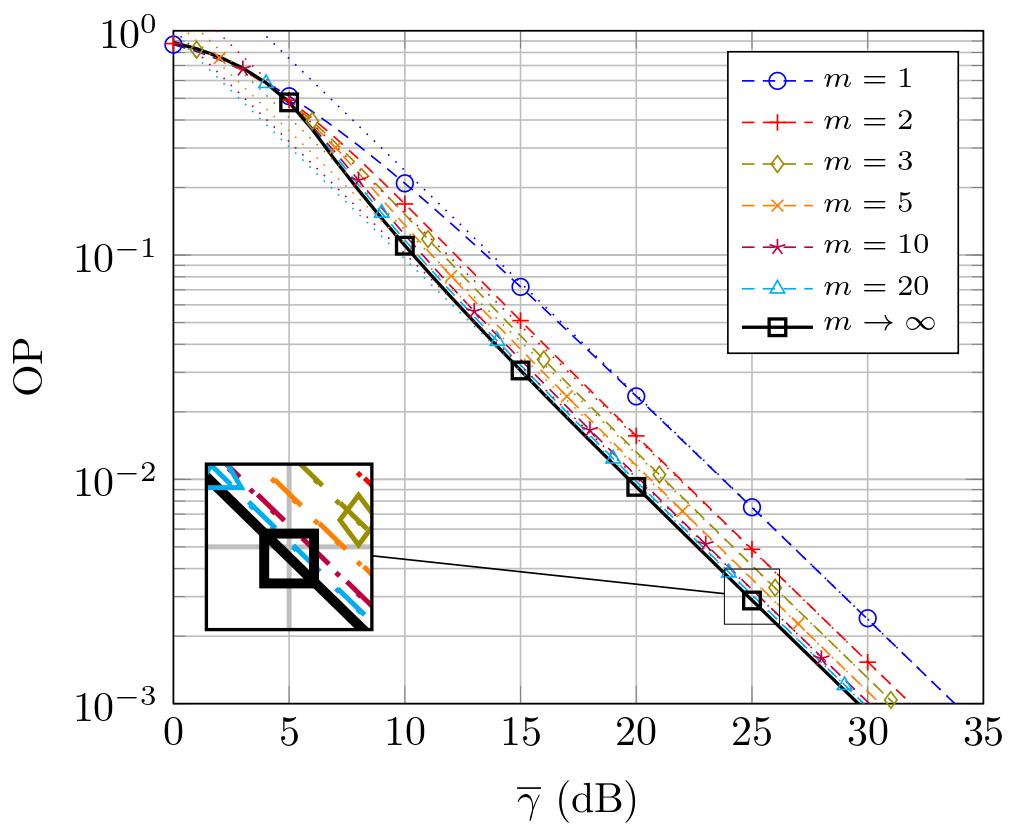}
\centering
\caption{\ac{OP} plot. Parameter values are $K=1$ and $\gamma_{\rm th}=3$dB. Dashed lines correspond to \ac{fdRLoS} fading. The case with $m\rightarrow\infty$ corresponds to \ac{dRLoS} fading \cite[eq. (21)]{Lopez2018}. Dotted lines correspond to the asymptotic \ac{OP} values in \eqref{aOP}. Markers correspond to \ac{MC} simulations.} 
\label{Fig3}
\end{figure}

We see that the \ac{OP} in \eqref{aOP} decays with a diversity order of 1, while the parameter $a$ acts as power offset or coding gain as defined in \cite{Wang2003}. Interestingly, such asymptotic expression is not valid for the specific case of $K=0$ as $\lim_{K\rightarrow 0} {\mathrm U}\left(m,1,K/m\right)
\rightarrow\infty$. This corresponds to the \ac{dR} or Rayleigh product channel, and for which the assumptions taken in \cite{Wang2003} cease to hold, i.e., the \ac{OP} for the \emph{pure} \ac{dR} case does not behave as ${\rm OP}_{\overline\gamma\rightarrow\infty}=\tfrac{a}{t+1}\left(\tfrac{\gamma_{\rm th}}{\overline\gamma}\right)^t$ for any $t>0$. In Fig. \ref{Fig3}, \ac{OP} is depicted for $K=1$, $\gamma_{\rm th}=3$dB  and for different values of $m$. As expected, the system performance improves as the \ac{LoS} fluctuation is reduced, i.e. the \ac{OP} decreases for higher values of $m$. Observe that for moderate values of $m$ the \ac{OP} behavior is practically indistinguishable from that of the deterministic \ac{LoS} (see the zoomed box in the figure). This reveals that the \ac{fdRLoS} model quickly converges as  $m$ grows to the baseline dRLoS starting model, a fact that is not so evident from the PDF curves in Fig.\ref{Fig1}.

\color{black}

\begin{figure}[t]
\centering
\includegraphics[width=.78\columnwidth]{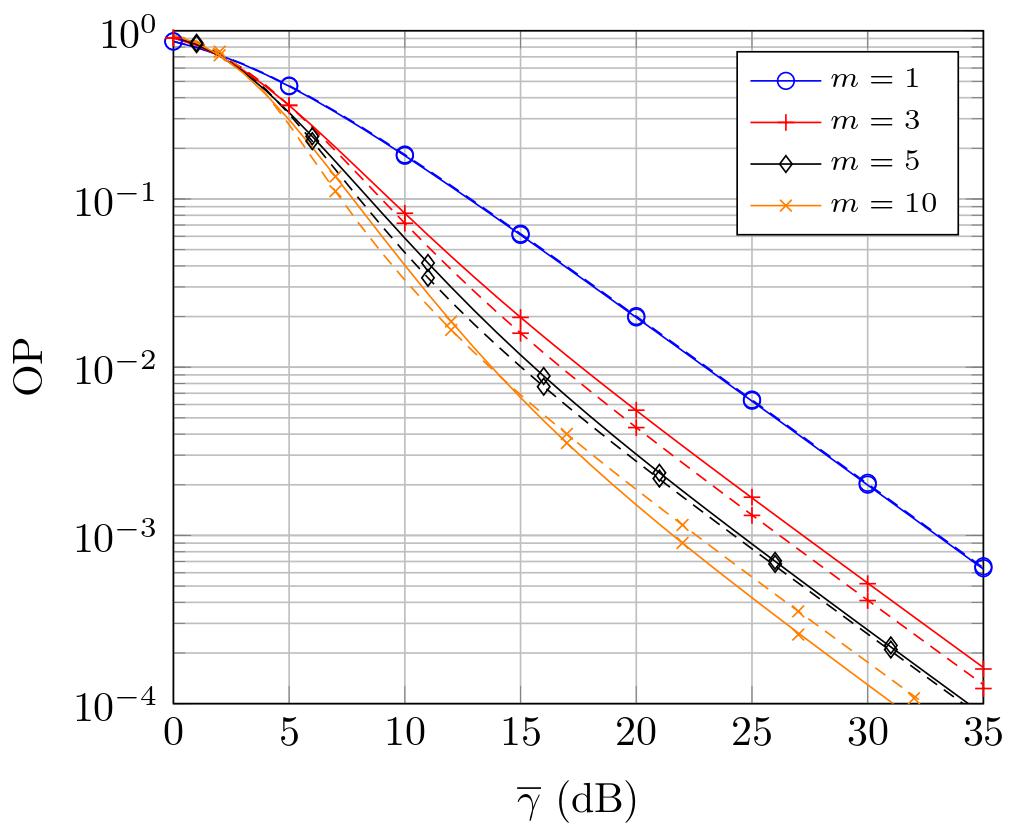}
\centering
\caption{OP comparison between \ac{fdRLoS} (dashed lines) and Rician shadowed (solid lines) fading models. \ac{OP} vs $\overline\gamma$ (dB), for different values of $m$. Parameter values are $K=6$ and $\gamma_{\rm th}=3$dB. Markers correspond to \ac{MC} simulations.} 
\label{Fig4}
\end{figure}

\begin{figure}[t]
\centering
\includegraphics[width=.78\columnwidth]{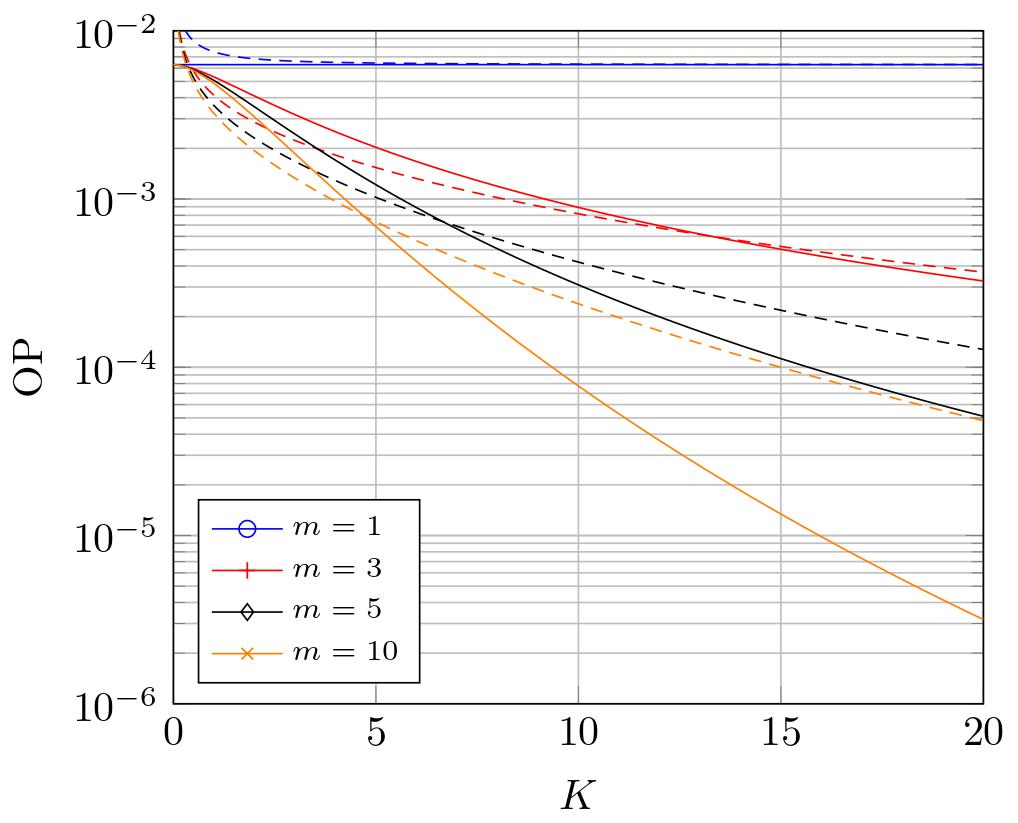}
\centering
\caption{OP comparison between \ac{fdRLoS} (dashed lines) and Rician shadowed (solid lines) fading models. \ac{OP} vs $K$ for different values of $m$. Parameter values are $\overline\gamma=25$dB and $\gamma_{\rm th}=3$dB.} 
\label{Fig5}
\end{figure}

Next, a comparison between the \ac{fdRLoS} (fluctuating \ac{LoS} and double-Rayleigh) and the Rician shadowed (fluctuating \ac{LoS} and Rayleigh) models is addressed. In the absence of \ac{LoS} component, it is known that a double-Rayleigh diffuse component leads to a worsening of the system performance as compared to a single-Rayleigh one. Surprisingly, this behavior may change in the presence of a \ac{LoS} component: in Fig. \ref{Fig4} the \ac{OP} as a function of $\overline \gamma$ is evaluated for the \ac{fdRLoS} (dashed lines) and Rician shadowed (solid lines) fading models with parameters $K=6$, $\gamma_{\rm th}=3$dB and different values of $m$. See that for $m=\{3,5,10\}$ there is a range of $\overline \gamma$ for which the OP of the \ac{fdRLoS} model falls below that of the Rician shadowed. This counterintuitive behavior is better reflected in Fig.\ref{Fig5} where the OP has been plotted as a function of $K$ with fixed parameters $\gamma_{\rm th}=3$dB and $\overline\gamma=25$dB. First, notice that for $K=0$ (no \ac{LoS}) the Rician shadowed model effectively yields a lower OP than the \ac{fdRLoS} model in all instances. As $K$ grows this relation is reverted in a range of $K$ whose size is dependent on $m$. In particular, see that this range is wider for moderate values of $m$, i.e. it is null for $m=1$, grows for $m=3$ and decreases for $m=5$ and $m=10$. 


 %
 %
\color{black}

\vspace{-2mm}
\section{Conclusion}
We introduced and characterized a new family of fading models consisting on a fluctuating \ac{LoS} component with \ac{dR} diffuse component. This model generalizes a class of fading models that arise fom the \ac{SOSF} model, and also exhibits a smoother behavior than its deterministic \ac{LoS} counterpart, thus avoiding one of the reported limitations of the \ac{dRLoS} model. We confirmed that in the presence of a moderate \ac{LoS} component, the performance of wireless communication systems operating under \ac{dR}-diffusely scattered fading can be better than for the single-Rayleigh case. This does not happen when the \ac{LoS} magnitudes are either small or large.

%
%
%
\vspace{-2mm}
\bibliographystyle{ieeetr}
\bibliography{Manuscript}

\begin{thebibliography}{10}

\bibitem{Zochmann2019}
E.~Z{\"o}chmann, S.~Caban, C.~F. Mecklenbr{\"a}uker, S.~Pratschner, M.~Lerch,
  S.~Schwarz, and M.~Rupp, ``{Better than Rician: modelling millimetre wave
  channels as two-wave with diffuse power},'' {\em EURASIP J WIREL COMM},
  vol.~2019, no.~1, p.~21, 2019.

\bibitem{Reig2019}
J.~{Reig}, V.~M. {Rodrigo Pe{\~n}arrocha}, L.~{Rubio}, M.~T.
  {Mart{\'\i}nez-Ingl{\'e}s}, and J.~M. {Molina-Garc{\'\i}a-Pardo}, ``The
  folded normal distribution: A new model for the small-scale fading in
  line-of-sight (los) condition,'' {\em IEEE Access}, vol.~7, pp.~77328--77339,
  2019.

\bibitem{Rice1945}
S.~O. {Rice}, ``Mathematical analysis of random noise,'' {\em Bell Labs Tech.
  J}, vol.~24, no.~1, pp.~46--156, 1945.

\bibitem{Andersen2002}
J.~B. Andersen, ``Statistical distributions in mobile communications using
  multiple scattering,'' in {\em Proc. 27th URSI General Assembly}, pp.~1--4,
  2002.

\bibitem{Salo2006}
J.~Salo, H.~M. El-Sallabi, and P.~Vainikainen, ``{Statistical Analysis of the
  Multiple Scattering Radio Channel},'' {\em IEEE Trans. Antennas Propag.},
  vol.~54, pp.~3114--3124, Nov 2006.

\bibitem{Lopez2018}
J.~Lopez-Fernandez and F.~J. Lopez-Martinez, ``{Statistical Characterization of
  Second-Order Scattering Fading Channels},'' {\em IEEE Trans. Veh. Technol.},
  vol.~67, pp.~11345--11353, Dec 2018.

\bibitem{Devineni2019}
J.~K. Devineni and H.~S. Dhillon, ``{Ambient Backscatter Systems: Exact Average
  Bit Error Rate Under Fading Channels},'' {\em IEEE Trans. Green Commun.
  Netw.}, vol.~3, no.~1, pp.~11--25, 2019.

\bibitem{Andrews1985}
L.~C. Andrews and R.~L. Phillips, ``{I--K distribution as a universal
  propagation model of laser beams in atmospheric turbulence},'' {\em J. Opt.
  Soc. Am. A}, vol.~2, pp.~160--163, Feb 1985.

\bibitem{Nikolaidis2018}
V.~Nikolaidis, N.~Moraitis, P.~S. Bithas, and A.~G. Kanatas, ``{Multiple
  Scattering Modeling for Dual-Polarized MIMO Land Mobile Satellite
  Channels},'' {\em IEEE Trans. Antennas Propag.}, vol.~66, pp.~5657--5661, Oct
  2018.

\bibitem{Abdi2003}
A.~{Abdi}, W.~C. {Lau}, M.~. {Alouini}, and M.~{Kaveh}, ``{A new simple model
  for land mobile satellite channels: first- and second-order statistics},''
  {\em IEEE Trans. Wireless Commun.}, vol.~2, no.~3, pp.~519--528, 2003.

\bibitem{Paris2010}
J.~F. {Paris}, ``{Closed-form expressions for Rician shadowed cumulative
  distribution function},'' {\em Electron. Lett.}, vol.~46, pp.~952 --953, June
  2010.

\bibitem{Churnside1989}
J.~H. Churnside and R.~G. Frehlich, ``{Experimental evaluation of log-normally
  modulated Rician and IK models of optical scintillation in the atmosphere},''
  {\em J. Opt. Soc. Am. A}, vol.~6, pp.~1760--1766, Nov 1989.

\bibitem{Simon2005}
M.~K. Simon and M.-S. Alouini, {\em Digital communication over fading
  channels}, vol.~95.
\newblock John Wiley \& Sons, 2005.

\bibitem{Paris2014}
J.~F. Paris, ``{Statistical Characterization of $\kappa$-$\mu$ Shadowed
  Fading},'' {\em IEEE Trans. Veh. Technol}, vol.~63, pp.~518--526, Feb 2014.

\bibitem{NIST}
``{\it NIST Digital Library of Mathematical Functions}.''
  http://dlmf.nist.gov/, Release 1.0.21 of 2018-12-15.
\newblock F.~W.~J. Olver, A.~B. {Olde Daalhuis}, D.~W. Lozier, B.~I. Schneider,
  R.~F. Boisvert, C.~W. Clark, B.~R. Miller and B.~V. Saunders, eds.

\bibitem{CHAUDHRY199499}
M.~Chaudhry and S.~Zubair, ``Generalized incomplete gamma functions with
  applications,'' {\em J Comput Appl Math}, vol.~55, no.~1, pp.~99 -- 123,
  1994.

\bibitem{Wang2003}
{Zhengdao Wang} and G.~B. {Giannakis}, ``A simple and general parameterization
  quantifying performance in fading channels,'' {\em IEEE Trans. Commun.},
  vol.~51, no.~8, pp.~1389--1398, 2003.

\end{thebibliography}


%
%


\end{document}